\documentclass{aa} 

\usepackage{graphicx}
\usepackage{natbib}
\bibpunct{(}{)}{;}{a}{}{,} 
\usepackage[varg]{txfonts}
\begin{document}

   \title{The complex environment of the bright carbon star TX~Psc as probed by spectro-astrometry\thanks{Based on observations made with ESO telescopes at La Silla Paranal Observatory under programme IDs 386.D-0091 and 091.D-0094.}}


   \author{
        J. Hron\inst{1} \and S. Uttenthaler\inst{1} \and B. Aringer\inst{1,2} \and D. Klotz\inst{1} \and 
        T. Lebzelter\inst{1} \and C. Paladini\inst{3} \and G. Wiedemann\inst{4}
          }

   \institute{
        University of Vienna, Department of Astrophysics, 
              T\"urkenschanzstrasse 17, A-1180 Vienna\\
              \email{josef.hron@univie.ac.at}
   \and Dipartimento di Fisica e Astronomia Galileo Galilei,
Universit\`a di Padova, Vicolo dell'Osservatorio 3, I-35122 Padova, Italy        \and Institut d'Astronomie et d'Astrophysique, Universit{\'e} Libre de Bruxelles, CP 226, Boulevard du Triomphe, 1050, Bruxelles, Belgium
         \and Hamburger Sternwarte, Gojenbergsweg 112, 21029 Hamburg, Germany
        }
   \date{Received ; accepted}

 
  \abstract
   {Stars on the asymptotic giant branch (AGB) show broad evidence of inhomogeneous atmospheres and circumstellar envelopes. These
        have been studied by a variety of methods on various angular scales. In this paper we explore the envelope of the well-studied carbon star TX\,Psc by the technique of spectro-astrometry.}
   {We explore the potential of this method for detecting asymmetries around AGB stars.}      
   {We obtained CRIRES observations of several CO $\Delta$v=1 lines near 4.6\,$\mu$m and HCN lines near 3\,$\mu$m in 2010 and 2013. These were then searched for spectro-astrometric signatures. For the interpretation of the results, we used simple simulated observations.}
   {Several lines show significant photocentre shifts with a clear dependence on position angle. In all cases, tilde-shaped signatures are found where the positive and negative shifts (at PA~0\degr) are associated with blue and weaker red 
components of the lines. The shifts can be modelled with a bright blob 70\,mas to 210\,mas south of the star with a flux of several percent of the photospheric flux. We estimate a lower limit of the blob temperature of 1000\,K. The blob may be related to a mass ejection as found for AGB stars or red supergiants. We also consider the scenario of a companion object.}
   {Although there is clear spectro-astrometric evidence of a rather prominent structure near TX~Psc, it does not seem to relate to the other evidence of asymmetries, so no definite explanation can be given. Our data thus underline the very complex structure of the environment of this star, but further observations that sample the angular scales out to a few hundred milli-arcseconds are needed to get a clearer picture.}

    \authorrunning{Hron et al.}          
    \titlerunning{The environment of TX\,Psc}

   {}\maketitle


\section{Introduction}

The presence of deviations from spherical symmetry in the photospheres and circumstellar envelopes of asymptotic giant branch (AGB) stars have become a well-established observational fact over the past decade. The asymmetries cover scales from fractions of a stellar radius (i.e.\ less than 1au) to thousands of au. Depending on the scale, several different mechanisms are causing the asymmetries ranging from temperature and density inhomogeneities due to convection cells \citep{chiavassa10} to the interaction between the stellar wind with a binary or the ISM (e.g.\ \citealp{maerckernature, jorissen11} and references therein). While the physical properties of the asymmetries are rather well known at those scales where imaging is possible (either directly or through radio-interferometry), the situation is much less favourable on the small scales where only indirect methods are available. On these small scales, observational evidence most frequently comes from optical interferometry, in particular from relative phases (either differential in wavelength or closure phase) or visibilities at high spatial frequencies \citep{paladini12,ragland06,chiavassa10, cruzalebes15}. Imaging with optical interferometry is continuously improving (e.g.\ \citealp{lebouquin09, berger12}; Paladini et al., in prep.), but it is mostly done at low-to-intermediate spectral resolutions (e.g.\ \citealp{ohnaka13antares}). This means that the kinematics associated with asymmetries are generally not acessible since the gas velocities in the atmospheres and envelopes of AGB stars are of the order of 10\,km\,s$^{-1}$ or less.  The L and M bands contain several lines that are sensitive to cool gas and to the temperature and density fluctuations expected for atmospheric inhomogeneities: the lines of the CO fundamental band around 4.6\,$\mu$m (e.g.\ \citealp{hinkleatlas95, ryde99}) or the HCN and C$_2$H$_2$ lines near 3.1\,$\mu$m (e.g.\ \citealp{jorgensen00, ridgway78}).  Unfortunately, no interferometer is currently operating in the L and M bands.

A method that gives access to the above lines at high spectral resolution and which efficiently allows angular resolutions to be achieved much better than that of the telescope is spectro-astrometry \citep{bailey98}. By measuring the precise position of the photocentre as a function of wavelength (so-called position spectra) and for different position angles~(PA) of the spectrograph slit, deviations from circular symmetry can be detected. The accuracy basically depends on the S/N and  the width of the PSF (e.g.\ \citealt{voigt09a}). Although it is an indirect way of mapping, this technique has been successfully used to investigate pre-main sequence binaries and protoplanetary disks (e.g.\ \citealp{bailey98, pontoppidan08}). A first application to cool giants was presented in \citet{voigt09a} and \citet{voigt09b}. In this paper we extend this work and present a spectro-astrometric study of the bright carbon star \object{TX Psc}.

In Section~\ref{target} we summarize previous results for our target, and Section~\ref{obs-reduction} describes the observations and the data reduction. Section~\ref{results} gives the results for the CO and HCN lines, Section~\ref{discussion} interprets the results in terms of different scenarios, and Section~\ref{conclusions} gives some conclusions.  


\section{The target}\label{target}

TX~Psc is among the brightest and closest carbon-rich AGB stars. It is classified as an irregular variable in the General Catalogue of Variable stars \citep{samus09}. However, an average period of 224\,d was detected in the photometry, and spectroscopic monitoring shows significant and regular velocity variations on a time scale of 450\,d \citep{jorissen11}. Thus the object is very likely a misclassified semiregular variable. The fundamental stellar parameters were derived by several authors, most recently in \citet{klotz13} using ISO/SWS spectra, VLTI/MIDI interferometric data, and hydrostatic model atmospheres. The good agreement between hydrostatic models and observations and the small variability indicate that the star's atmosphere is almost free of dust and much less dynamic than the ones of mira variables. 

Several observations,however,  point to a more complex structure of the upper atmosphere and envelope. At the largest scales, \citet{jorissen11} detected an almost circular shell in the far-IR with a radius of about 17\arcsec. Interaction of the stellar wind with the surrounding ISM could only have caused this shell if the mass loss rate had been much higher in the past than the current value of about $10^{-7}M_{\sun}$yr$^{-1}$ \citep{olofsson93}. Evidence of a 10\,\arcsec\ bipolar structure in the NE-SW direction comes from mm-CO observations \citep{heske89b}, though an alternative interpretation would be that of a clumpy wind. An axisymmetry in the same direction is also seen in the Herschel far-IR data of \citeauthor{jorissen11} on the arcmin scale, and this direction coincides with the proper motion of the star. The presence of the circular shell between these asymmetries makes it very unlikely that the asymmetries have the same physical origin. 

\citet{cruzalebes98} were able to resolve the envelope surrounding TX\,Psc in the K band with adaptive optics and detected a clump of size $<$0.25\arcsec at a separation from the star $\sim$0.35\arcsec\ and towards SW, i.e.\ roughly the same direction as the mm-CO asymmetry. The flux of this blob is only about 2\% of the central flux. Using the lunar-occultation technique, \citet{richichi95} found different types of asymmetries on scales of a few mas between the V and L bands. The K-band data show notable differences in the brightness profiles between different position angles (PA), while the L-band profile is double peaked along the NE-SW direction, thus supporting the mm-CO observations. Non-zero IOTA closure phases observed by \citet{ragland06} for TX~Psc can be qualitatively reproduced by an unresolved spot at the edge of the stellar disk, i.e.\ at a few \,mas distance. Most recently, \citet{cruzalebes13} presented evidence that there is a departure from centrosymmetry based on VLTI/AMBER data. On the other hand, MIDI interferometry carried out by \citet{klotz13} was found to be consistent with a spherical structure of the star and its circumstellar envelope. 

There is also purely spectroscopic evidence supporting a non-homogeneous upper atmosphere. Chromospheric emission in the UV has been found for carbon stars since the 1960s (see review by \citealp{luttermoser00}), and the suggestion that these chromospheres are non-homogeneous was made by \citet{jorgensen91} on the basis of a comparison between chromospheric models and IR spectra. The detection of IR fine-structure emission lines in ISO spectra of TX~Psc was interpreted by \citet{aoki98b} as evidence of a 1000\,K layer above the photosphere. Finally, the lack of a deep absorption around 14\,$\mu$m in the ISO-SWS spectrum of TX~Psc was interpreted by \citet{jorgensen00} as due to clumps of $\sim~$500\,K with a filling factor of about 10\%. In spite of the variety of available observations, no clear picture exists about the presence of clumps or even a bipolar structure around TX~Psc. This is partly because the data are distributed over two decades and partly because they do not combine high spectral and angular resolution. The latter condition is fulfilled by the data presented in the following. 

We want to emphasize at this point that this work concentrates on the spectro-astrometric evidence for asymmetries in the M and L bands and does not aim at a complete spectroscopic analysis in terms of fundamental stellar parameters. This will be the subject of future work.



\section{Observations and data reduction}
\label{obs-reduction}

\begin{figure}
\centering
\includegraphics*[width=7.cm]{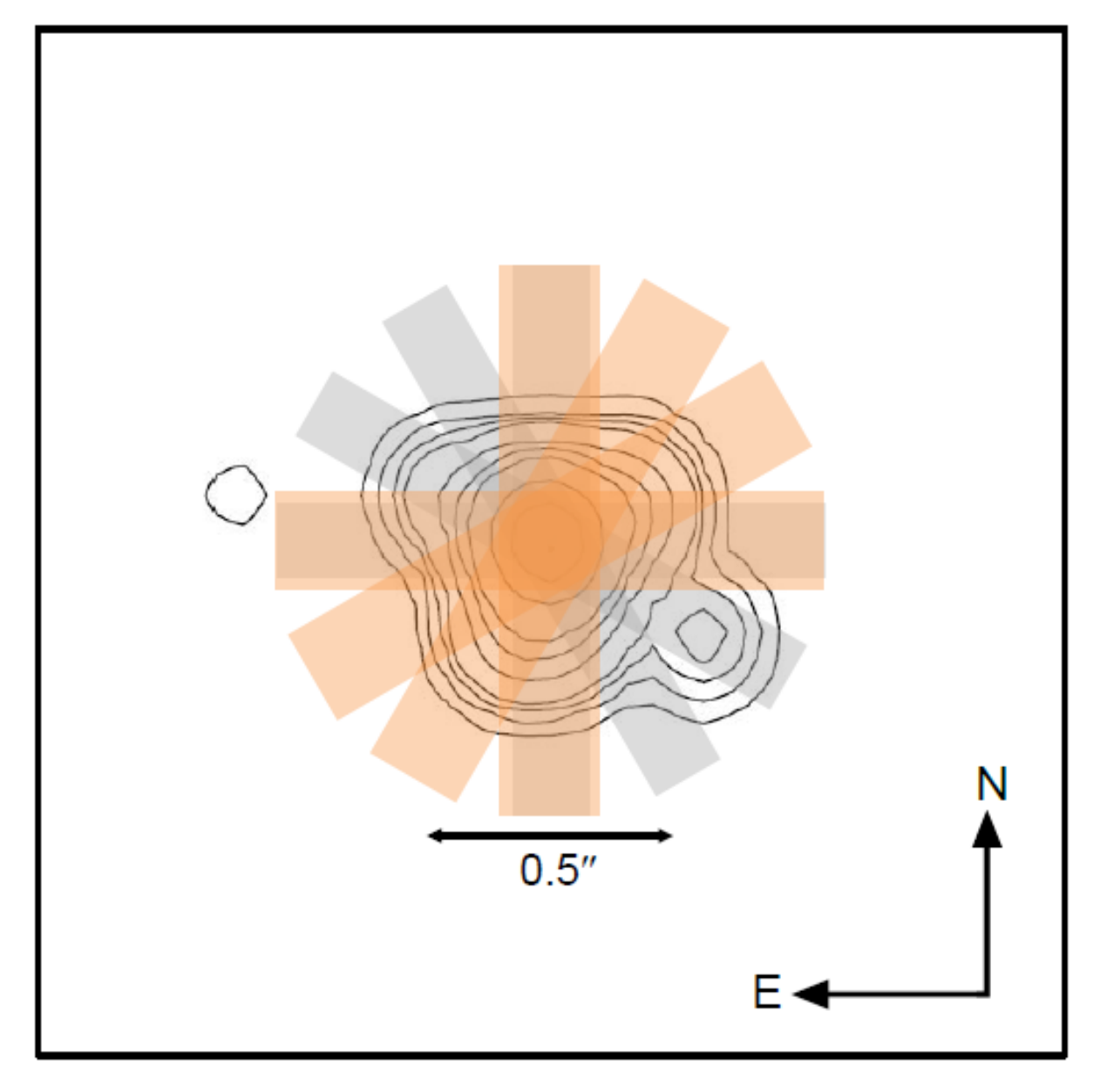}
\caption{\label{nacofig}Reconstructed map of TX~Psc in the K band from \citet{cruzalebes98} with the
CRIRES slit positions used superposed. The FWHM of the NACO beam was 0.15\arcsec.The width of the rectangles roughly corresponds to the slit widths used.
The NE/SW quadrants were observed in 2010 (grey rectangles), the 
NW/SE quadrants in 2013 (orange). PAs of 0\degr\ and 90\degr\ were observed in both runs. The outermost four contours 
correspond to 0.1, 0.3, 0.6, and 1\% of the central flux. }\end{figure}

\object{TX Psc} was observed in 2010 and 2013 with the CRIRES spectrograph \citep{kaeuflcriresspie} at the VLT. The observations in 2010 were carried out in the night of October~27 under very good conditions (seeing 0.6\arcsec). Four different wavelength settings in the K, L, and M bands were used, and the slit width was 0.15\arcsec. The observations in 2013 were performed on September 14 under reasonable conditions (seeing around 1\arcsec, cloudy). The slit width was 0.2\arcsec, and one pixel corresponds to 86\,mas. Three wavelength settings in the K, L, and M bands were used, with the settings in the M band identical to the one of 2010. The M-band data and the L-band observations of 2010 contain the best diagnostic lines and are used in this paper, and the remaining data are the subject of a forthcoming paper. The brightness of the target required the use of a windowed detector read-out mode
\footnote{http://www.eso.org/sci/facilities/paranal/instruments/crires/doc.html} where only the central two of the four detectors are read out, thus allowing the shortest detector integration times. The 2010 and 2013 data in the M band cover wavelengths between 4.594\,$\mu$m and 4.6495\,$\mu$m and contain several CO $\Delta$v=1 lines. In addition, we also discuss the results in the wavelength range from 2.985\,$\mu$m to 3.017\,$\mu$m obtained in 2010. This range is located in the centre of the prominent HCN(001) band (e.g.\ \citealp{hron_rscl}). For all observations presented here, the AO system of CRIRES was used. Nodding along the slit was used to correct for the background.  

\begin{figure}
\centering
\includegraphics*[width=9.cm]{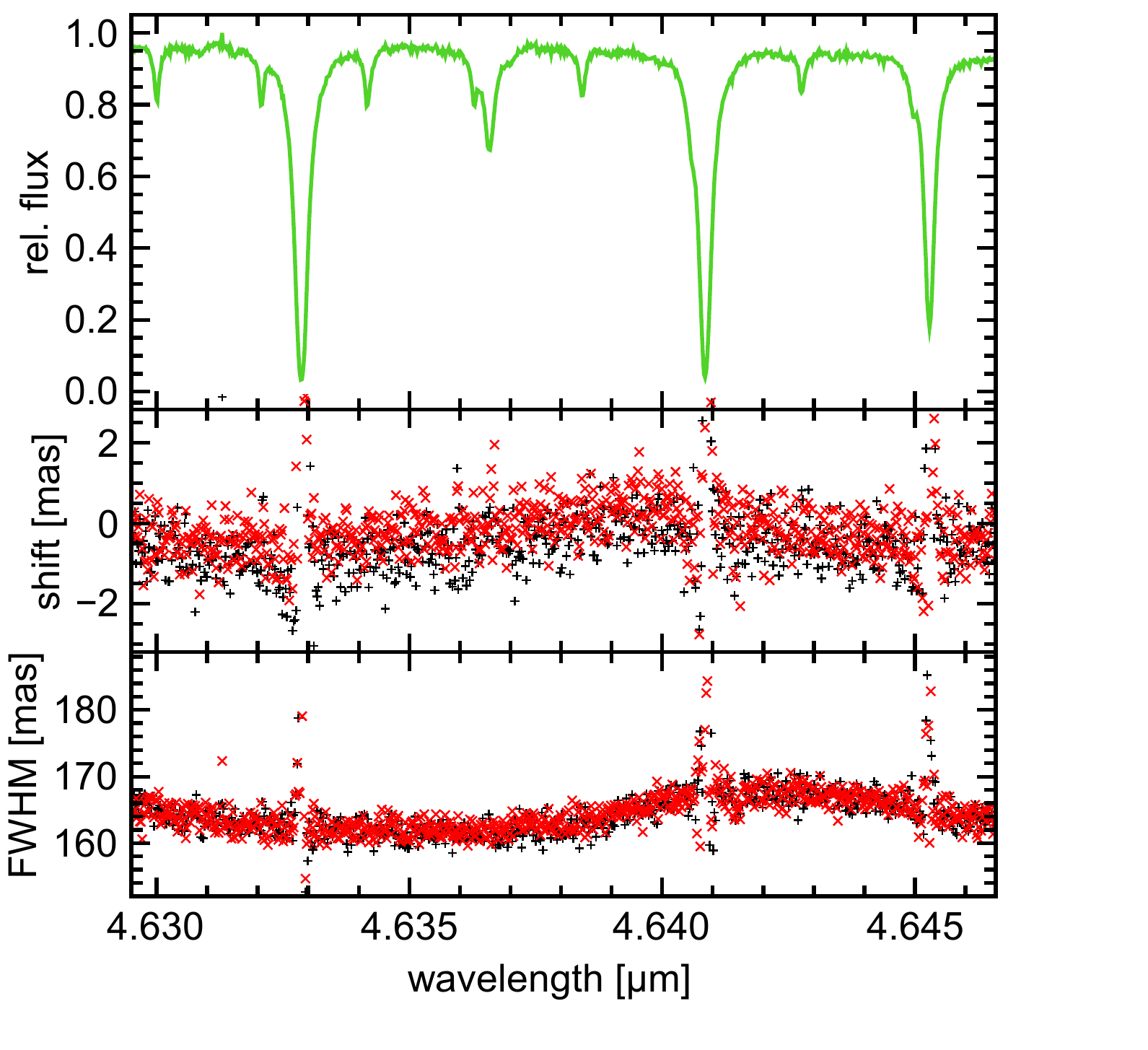}
\caption{\label{hipp-posspectrum}Example 2010 data for the early type standard \object{HIP~116928} in the M band. The uppermost panel shows the normalized flux spectrum. The middle and lower panels are the position and FWHM spectra resulting from the Gauss fits, respectively. For the position spectra, a global trend was subtracted (see text). The plus signs and crosses correspond to the data at PAs of 0\degr\ and 180\degr, respectively.}
\end{figure}

\begin{figure}
\centering
\includegraphics*[width=9.cm]{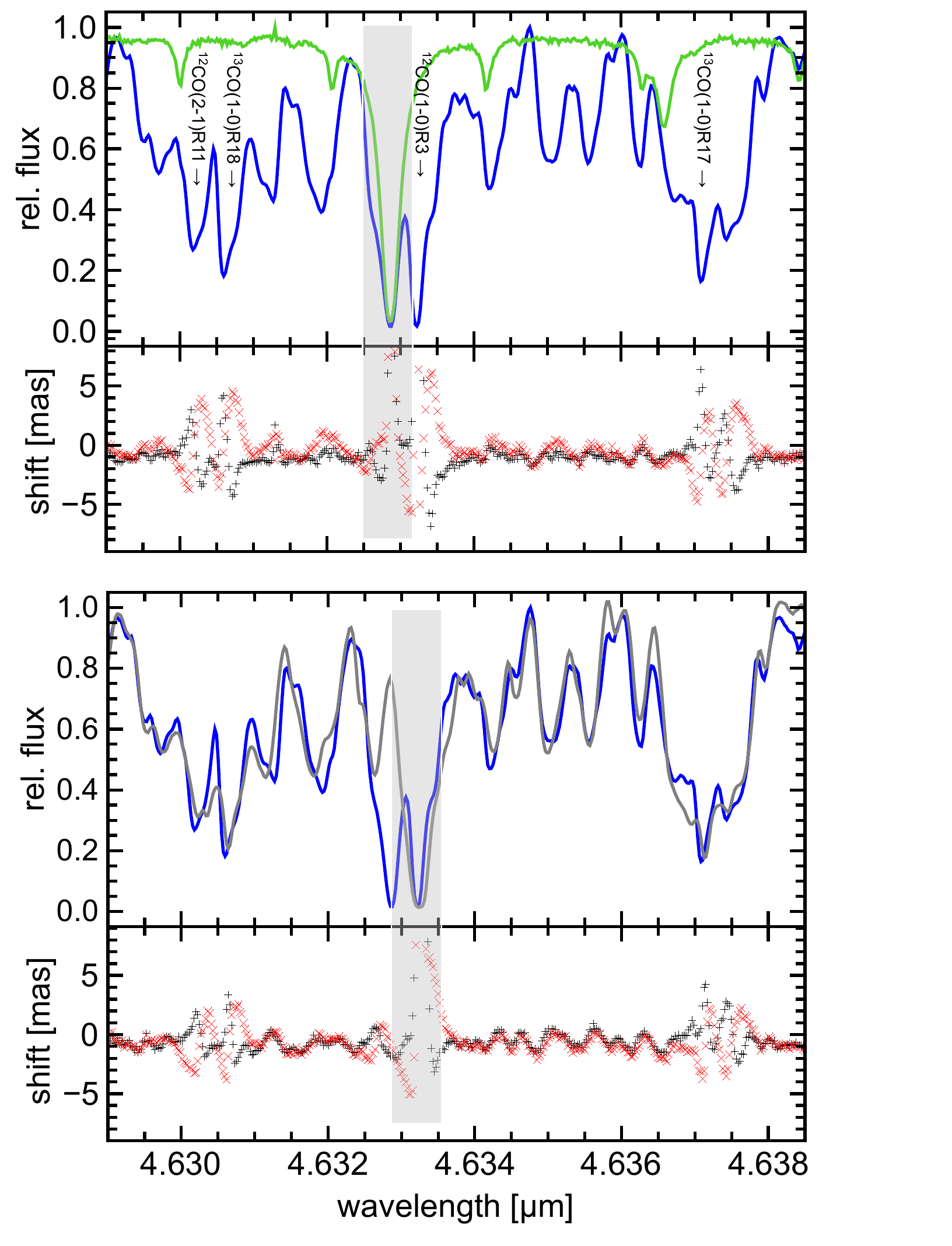}
\caption{\label{posspectra1}Flux and position spectra for 2010 and 2013 near 4.63\,$\mu$m. In the upper panel, the blue curve is the flux spectrum of TX~Psc in 2010, the green curve the spectrum of the early type standard \object{HIP~116928}. In the lower panel, the blue curve is the flux spectrum of TX~Psc in 2010, and the grey curve is the spectrum in 2013. The CO lines used in this analysis are identified, the grey rectangles indicate regions affected by telluric absorption. All spectra have been shifted to the wavelength scale of 2013, therefore the telluric absorptions are red-shifted by about 0.0005$\mu$m in the lower panels. The plus signs and crosses correspond to the position spectra at PAs of 0\degr\ and 180\degr, respectively, for the 2010 (upper panel) and 2013 (lower panel) data. Note the tilde-shaped profiles in the position spectra of the stellar CO lines and the strong photocentre deviation in the telluric lines.
The behaviour of the position spectra in the marked lines is almost identical at the two epochs.}
\end{figure}

To constrain any detected asymmetry in its azimuthal extent, observations were repeated at different slit PA: 0\degr, 30\degr, 60\degr\, and 90\degr\ in 2010 and 0\degr, 90\degr, 120\degr\, and 150\degr\ in 2013. The main concern of spectro-astrometry are artefacts caused by the shape of the PSF and internal instrumental effects. A simple and quite effective way to detect these artefacts is to obtain spectra at PA and PA+180\degr\  \citep{brannigan06} since real signatures should show up with opposite astrometric signs. This procedure was also used for our observations. Figure \ref{nacofig} shows
the slit positions superposed on the K-band map of \citet{cruzalebes98}. To identify telluric lines and assess the accuracy
of the spectro-astrometry, early-type standard stars were observed with the same set-up and PAs right after or before the observations of TX~Psc.
Data were reduced with the CRIRES pipeline 
\footnote{http://www.eso.org/sci/facilities/paranal/instruments/crires/tools.html}, 
and the achieved S/N was above 100 in the (pseudo)continuum regions. Centroid positions were derived from Gauss fits perpendicular to the dispersion direction. The profiles perpendicular to the dispersion direction showed a weak asymmetry towards increasing pixel numbers that is due to optical aberrations in the spectrograph (K\"aufl, private communication). In spite of this asymmetry, we decided to use Gauss fits for the following reasons: (i)~the asymmetry is largely independent of target, wavelength, time, and PA; (ii)~we measure the photocentre-shifts always relative to neighbouring wavelengths, i.e.\ any systematic offset does not influence the results (see also \citealp{voigt09a}); (iii)~the profiles are defined over only 12~pixels, thus making a fit with a more
elaborate function problematic. To minimize artefacts in the spectro-astrometry, the individual nodding positions were measured separately (i.e. not combined), and no correction for telluric absorption lines was applied. The mean FWHM of the Gauss fits along the slit as measured from the standard stars and our target was always between 2 and 2.4 pixels, corresponding to 0.17\arcsec\ and 0.2\arcsec\, respectively (Fig.~\ref{hipp-posspectrum}). This is close to the diffraction limit at these wavelengths, indicating the good adaptive optics performance of CRIRES.

The position spectra generally showed global trends with wavelength due to spectral curvature and a slight misalignment of the dispersion direction and the detector rows. These trends were removed by subtracting polynomials of second order. After this correction, the variations in the centroid positions were much less than 0.1\,pixels (Fig.~\ref{hipp-posspectrum}). The
agreement between opposite PA was always much better than this except near strong telluric lines and in cases of real spectro-astrometric signatures. Position and flux spectra of the regions with spectro-astrometric signals are shown in Figs.~\ref{posspectra1}, \ref{posspectra2}, \ref{posspectra3}, \ref{3muposspectra1}, \ref{3muposspectra2}, and \ref{3muposspectra3}.
\footnote{The appendix is only available in electronic form at http://www.aanda.org}

If opposite PA showed opposite deviations in the position spectrum for both nodding beams and if there was no notable contamination from telluric lines, this was considered as a real spectro-astrometric signal. As in most cases the photocentre shifts of opposite PA were of equal size (but opposite sign), so half the total centroid difference between opposite PA was then defined as the actual photocentre shift. These shifts were always accompanied by an increase in FWHM. In the case of real shifts, this amounted to less than 0.01\arcsec. Since this increase was the same for PA and PA+180\degr\, it was not used as a criterion to define a real photocentre shift.

Fundamental band CO lines were identified based on the atlas of \citet{hinkleatlas95}. For the HCN lines near 3$\mu$m, we
used a model atmosphere with $T_{\rm{eff}}=3100$\,K, $Z/Z_{\sun} = 1$, $log$(g[cm\,s$^{-2}])=-0.5$, $M = 1\, M_{\sun}$, and C/O$ = 1.05$ \citep{klotz13}. For this model we calculated a synthetic spectrum containing only HCN \citep{aringer09}. For the heliocentric stellar velocity, we adopted 10\,km\,s$^{-1}$, in between the velocities obtained in the visual and for CO\,($J=1-0$) lines (\citealt{jorissen11} and references therein).
The exact choice of this value is, however, not critical for our analysis.
 

\section{Results}\label{results}

\subsection{The CO $\Delta$v=1 lines}
\label{res.CO}

\begin{figure}
\centering
\includegraphics*[width=9.cm]{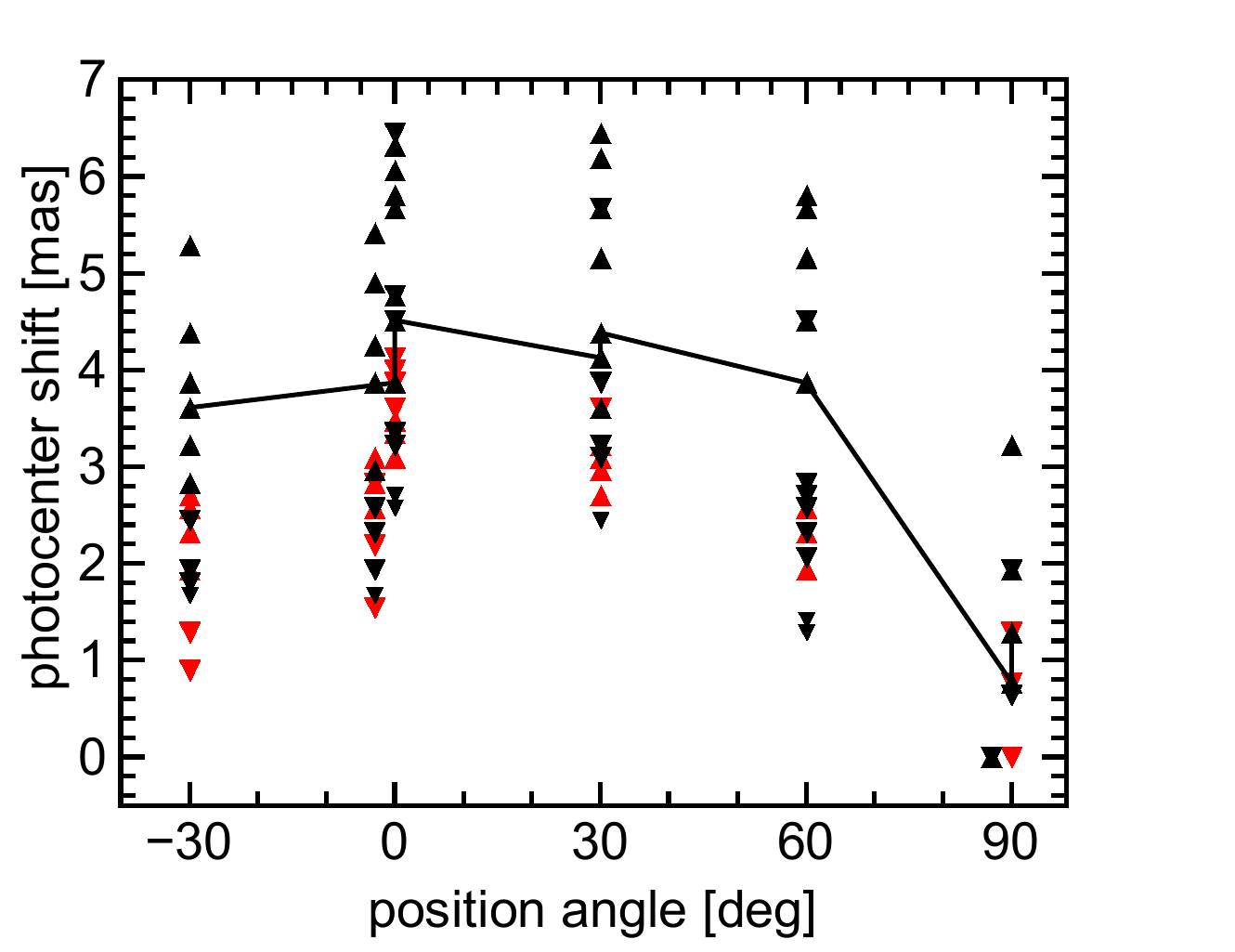}
\caption{\label{allshifts}Photocentre shifts of all used CO lines versus PA.  Black and red symbols refer to $^{13}$CO(1$-$0) and $^{12}$CO(2$-$1) lines, respectively. Up and down triangles give the measurements for the blue and red line components, respectively. For each line and PA, two measurements are plotted corresponding to the two nodded spectra. The data for the $^{13}$CO(1$-$0)R21 line
are connected by a line to guide the eye. The PA$=$0\degr\ and 90\degr\ data of 2013 have been offset in PA by $-$3\degr\ for clarity. The PA$=-$60\degr\ data are not shown because their offset is not significantly different from zero.}
\end{figure}

We have detected signals for the $^{12}$CO(2$-$1)R9, 10, 11, and 14 and for the $^{13}$CO(1$-$0)R16, 17, 18, 20, and 21 lines with a clear dependence on PA. Except for the 
$^{13}$CO(1$-$0)R20 line, all lines showed a tilde-shaped signal, i.e.\ a positive (southern) shift for PA=0\degr\ followed by a negative shift at a slightly longer wavelength (see Figs.~\ref{posspectra1}, \ref{posspectra2}, and \ref{posspectra3}). As can be noted in the figures, all the lines with a spectro-astrometric signal show asymmetric profiles with a second component at redder wavelengths. This red component is associated with the red part of the tilde-shaped signal (see below). The individual photocentre shift measurements are shown in Fig.~\ref{allshifts} as a function of PA. On the detector, south is in the direction of increasing pixel numbers. Thus the observed signs of the shifts for all PA are consistent with an asymmetry towards the south.

Some of the $^{12}$CO(1$-$0) lines also showed indications of spectro-astrometric signals, but it was not possible to measure these reliably because of the contamination by telluric absorption. Three $^{12}$CO(3$-$2) lines possibly show a weak shift but only at one PA and only in one direction. Given the highly uncertain signal, we do not consider these lines any further.

\begin{figure}
\centering
\includegraphics*[width=9.cm]{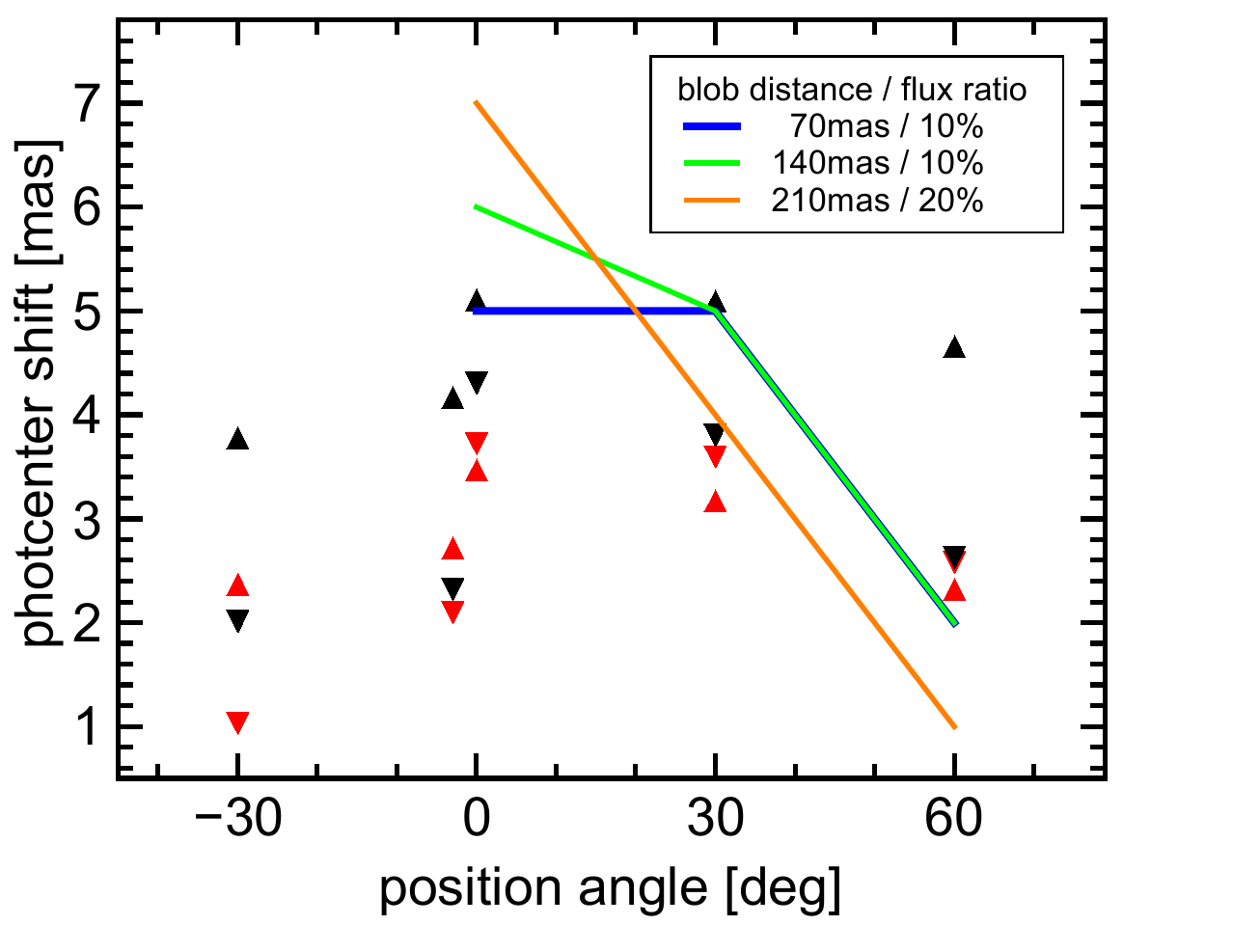}
\caption{\label{meanshifts}Mean photocentre shifts of all used CO lines versus PA. Colours and symbols as in Fig.~\ref{allshifts}. The PA$=$0\degr\ data of 2013 have been offset in PA by $-$3\degr\ for clarity, the PA$=$90\degr\ and $-$60\degr\ data are not shown because their offset is not significantly different from zero. The blue, green, and orange lines show the predictions from the simulations (see Sect.~\ref{discussion}) for three different combinations of blob distance and flux ratio.}
\end{figure}

The mean shifts for the different line groups and components are shown in Fig.~\ref{meanshifts} where the measurements from 2013 are slightly shifted in PA for clarity. The photocentre shifts for PA$=$90\degr\ in 2010 and 2013 are not significantly different from zero, so these data are not plotted. It is evident that the $^{12}$CO(2$-$1) lines show a weaker signal than the $^{13}$CO(1$-$0) lines and that the shifts in the blue line components are greater than those of the red components. Comparing the data from 2010 and 2013, the maximum shift at PA$\approx$0\degr\ indicated by the 2010 data alone is confirmed by the measurements of 2013; however, the shift found for the 2013 data is lower.

\begin{figure}
\centering
\includegraphics*[width=9.cm]{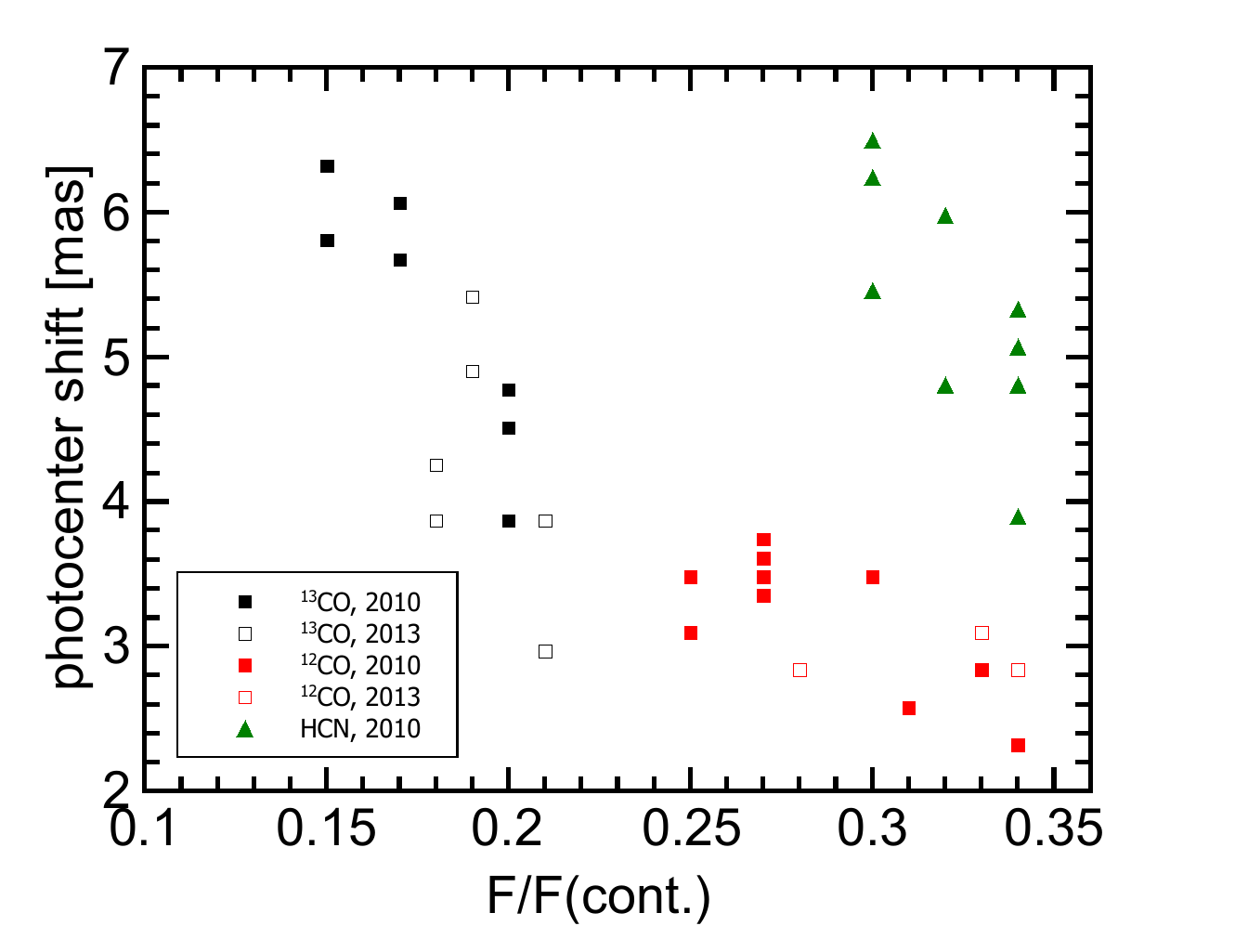}
\caption{\label{shifts-depth} Photocentre shifts of the blue components of all used CO and HCN lines for PA$=$0\degr\ versus line depth. Colours as in Fig.~\ref{allshifts} for the CO lines. Green triangles refer to the HCN lines.
Full and open squares denote the 2010 and 2013 CO data, respectively.
For each line, the shifts of both nodding positions are plotted.}
\end{figure}

The detected shifts are not correlated with the energy of the lower state term but with the depth of the line relative to an assumed flat continuum. Deeper lines show a higher shift. This is illustrated in Fig.~\ref{shifts-depth} where we have
only used the blue line components for which the depth can be reliably measured. The two nodding positions for each line are plotted separately. The difference in the shifts between the $^{12}$CO$(2-1)$ and $^{13}$CO$(1-0)$
is likely to be caused by the significantly different line depths, i.e. the flux from the source causing the
shift is probably constant in the region of the photospheric lines, but the photospheric flux in the cores of the $^{12}$CO$(2-1)$ lines is larger, thus reducing the photocentre shift. The lines in 2013 appear to be typically weaker by 10\% to 20\% than in 2010, which could explain the change in the shifts for a given PA from 2010 to 2013. Although the uncertainty in the true continuum could be responsible for the apparent difference in line depths between the 2010 and 2013 data, intrinsic temporal line variations cannot be excluded. Another possible influence on the shifts is the broader slit used in 2013, which might reduce the effects of asymmetries. Since the slit width and the FWHM of the spectra along the slit were comparable at both epochs, we expect this effect to be small. This is supported
by the simulations presented in Sect.~\ref{discussion}.

Finally we note that, owing to the changing weather conditions in 2013, some observations without adaptive optics
were also
performed. For none of these were photocentre shifts detected in accordance with the broader PSF. In summary, we are confident that the spectro-astrometric signals described above are real and not artefacts. 


   

\begin{figure}
\centering
\includegraphics*[width=9.cm]{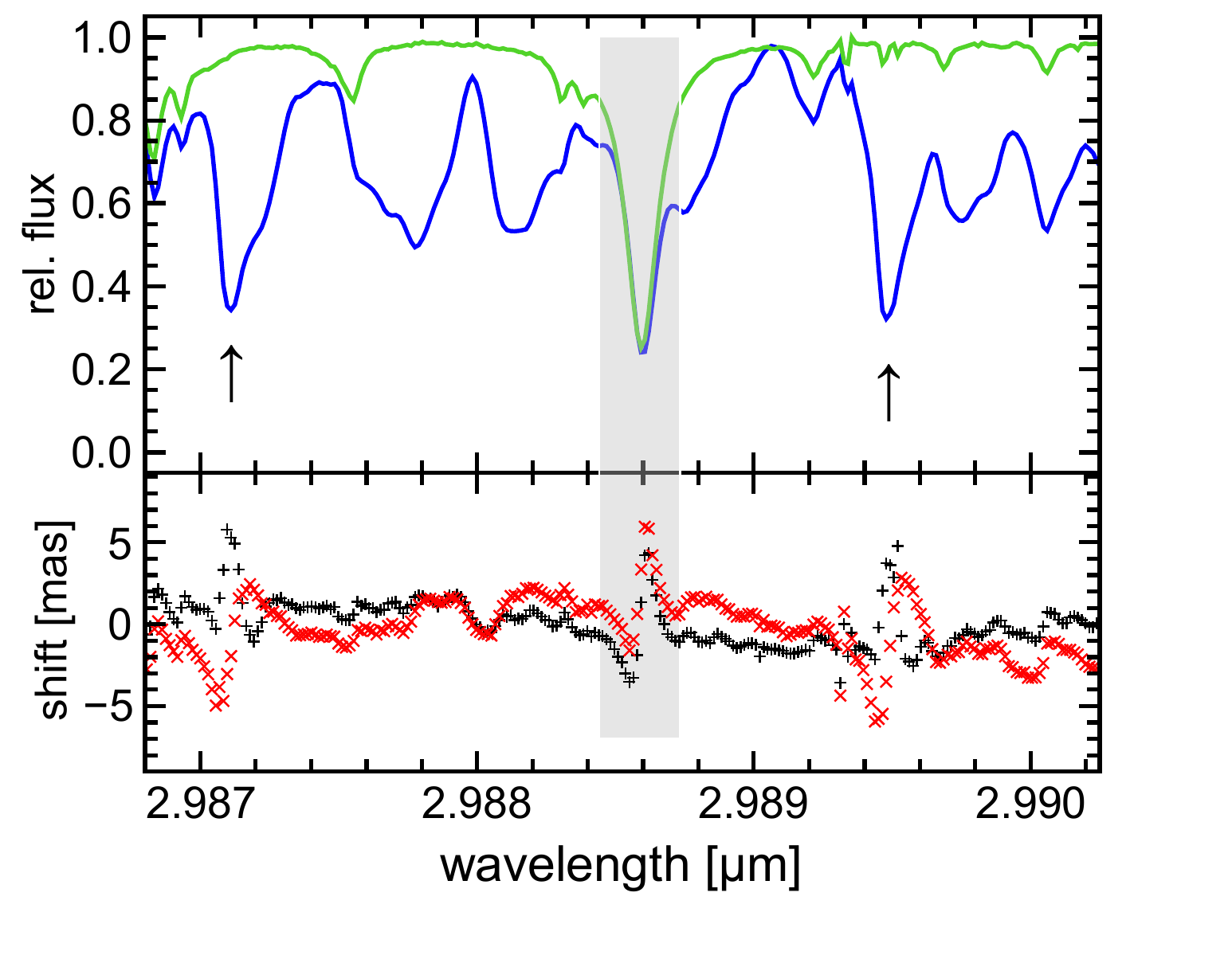}
\caption{\label{3muposspectra1}Flux and position spectra in 2010 near 2.98\,$\mu$m. The blue curve is the flux spectrum of TX~Psc, and the green curve is the spectrum of the early type standard \object{HIP~116928}. The arrows identify the HCN lines used in the analysis, the grey rectangles indicate regions affected by telluric absorption. 
The plus signs and crosses correspond to the position spectra at PAs of 0\degr\ and 180\degr, respectively. We note again the tilde-shaped profiles in the position spectra of the stellar HCN lines.}
\end{figure}

\subsection{The HCN lines}
\label{res.HCN}
The procedure for searching for spectro-astrometric signatures was the same as for the CO lines. The mean variations in the centroid positions with wavelength and the agreement between opposite PA were similar to what was found for the CO lines (better than 0.1\,pixels). We detected spectro-astrometric signatures in four HCN lines with a clear dependence on PA. Each line showed the same behaviour as found for CO, i.e.\ a tilde-shaped signal associated with the blue and red components
of the line (see Figs.\, \ref{3muposspectra1}, \ref{3muposspectra2}, \ref{3muposspectra3}).

\begin{figure}
\centering
\includegraphics*[width=9.cm]{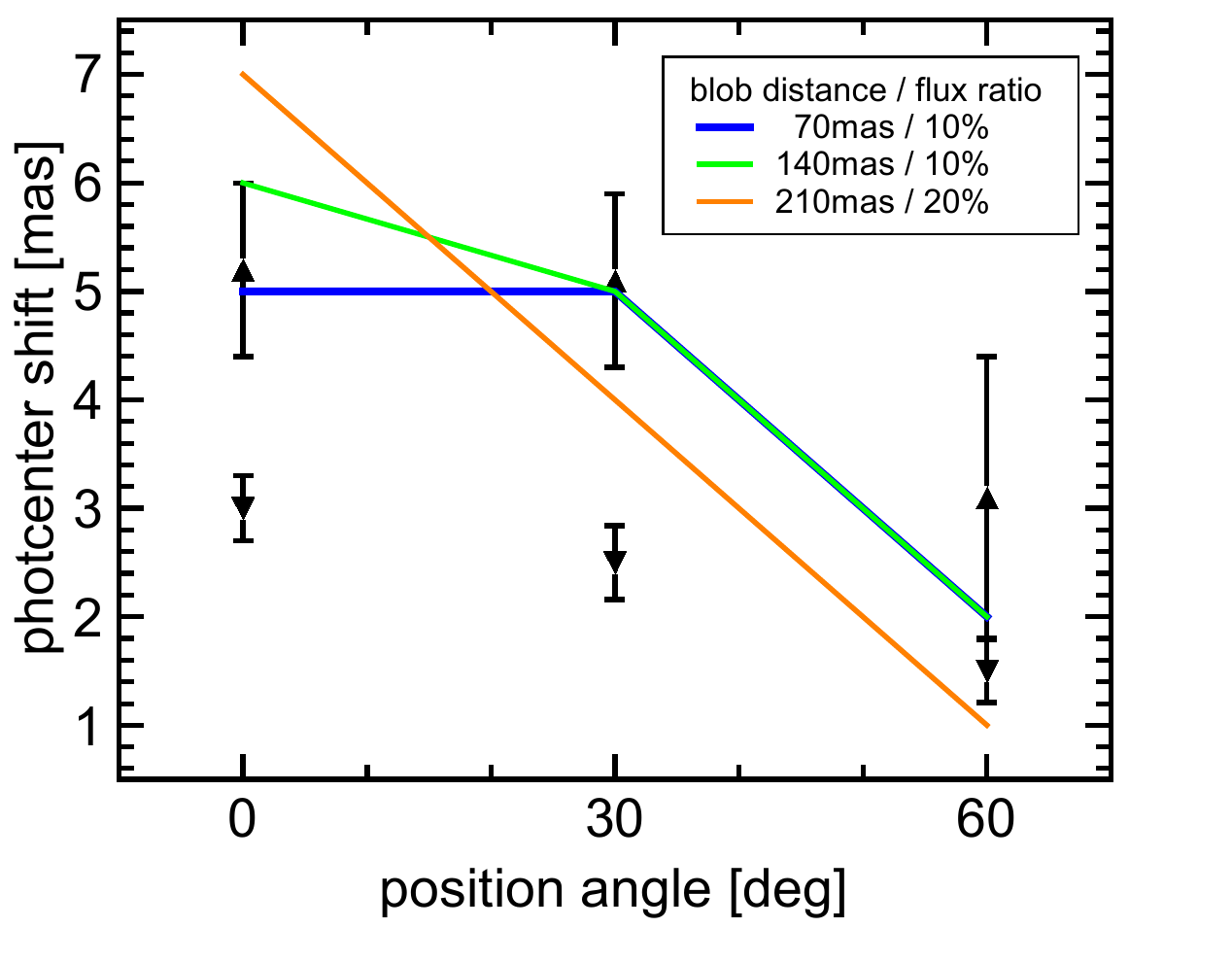}
\caption{\label{3mumeanshifts}Mean photocentre shifts oin all used HCN lines versus PA. The PA$=$90\degr\ data are not shown because their offset is not significantly different from zero. Up and down triangles give the measurements for the blue and red line component, respectively. The blue, green, and orange lines show the predictions from the simulations (see Sect.~\ref{discussion}) for three different combinations of blob distance and flux ratio.}
\end{figure}

The mean photocentre shifts are shown in Fig.~\ref{3mumeanshifts}. When comparing this with Fig.~\ref{meanshifts}, it is evident that the shifts in the blue components are similar to the ones of the $^{13}$CO$(1-0)$ lines. Although the red
components also show a decreasing shift with increasing PA, the shifts are less than for the $^{13}$CO$(1-0)$ lines. The maximum shift is again found for PA$\approx$0\degr, but the shifts at PA 60\degr\ are lower for the HCN lines. For the CO lines, we note a dependency of the photocentre shift on line depth (Fig.~\ref{shifts-depth}), although the HCN lines show much higher shifts at the same
line depth. The higher photocentre shifts in the HCN lines are therefore due to a wavelength-dependent property of the structure causing the photocentre shift. 
The difference between the blue and red components of the HCN lines is probably again caused by the smaller depth of
the red components, i.e.\ the larger photospheric flux at the wavelength of the red component.

\section{Discussion}
\label{discussion}
The observed tilde-shaped photocentre shifts found for the CO and HCN lines indicate an
asymmetry of the object towards the south (based on the orientation of the detector on the sky for PA 0\degr). The shape can be understood if the blue line components are from the stellar photosphere, and the positive shifts (for PA 0\degr) are caused by continuous emission from a bright blob south of the star. The weaker red line components would then be associated with red-shifted absorption in front of the blob. If this red-shifted absorption is due to a mass outflow (as expected from the mm-CO observations), the blob must be located on the far side of the star. The opposite photocentre shift at the wavelengths of the red line component is caused by the dominant continuum emission from the stellar photosphere. This scenario would also explain the lower shifts in the red components where the absorption is not as deep; i.e.,\ the flux from the stellar photosphere  is higher.

We note at this point that a dark or bright spot on the surface of the star would also produce a photocentre shift. However, the diameter of TX~Psc is only between 8\,mas and 11\,mas between the optical and 10$\mu$m \citep{klotz13}, and thus any photocentre shift by a spot could not be detected with our data.  A tilde-shaped signature could also be due to stellar rotation \citep{lesage2014}, but in this case one would just
see a rotationally broadened spectral line and not two line components. Finally, an elliptical instrumental PSF, as caused by astigmatism or imperfect active/adaptive optics, would also produce the observed tilde-shaped shifts \citep{voigt09a}. However, such an instrumental effect should not change sign for PA+180\degr, it would not lead to the red line components in the spectrum, and it would probably not be constant during the night. Furthermore, we observed other targets with similar spectra on both nights, and these did not show the tilde-shaped shifts. Therefore we do not consider any of these possible scenarios further.

\begin{figure}
\centering
\includegraphics*[width=9.cm]{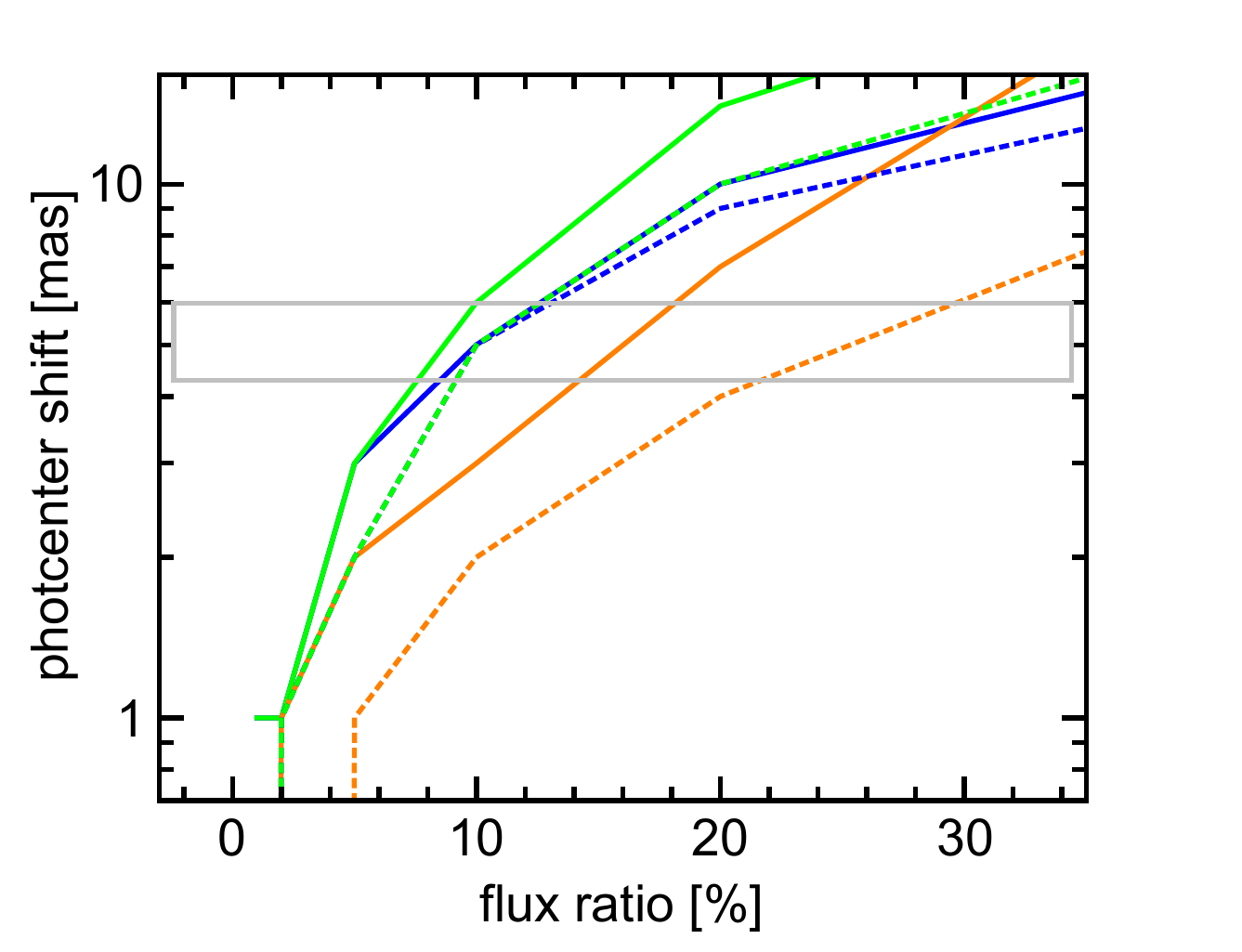}
\caption{\label{shift-simulations} Photocentre shifts as predicted by the superposition of two Gaussian sources with
different distance, flux ratio, and position angle (see text for details). The full and dashed lines correspond to PAs of
0\degr\ and 30\degr, respectively. The blue, green, and orange curves are for source distances of 70, 140, and 210\,mas, respectively. The full rectangle indicates the range (mean$\pm 1\sigma$) of shifts in the blue components of the $^{13}$CO(1$-$0) lines in 2010.}
\end{figure}

To get a rough estimate of the blob properties, we performed simple numerical simulations. We created artificial images consisting of two Gaussian sources of variable distance, PA, and flux ratio. For the FWHM we used the
PSF width as derived from the Gauss fits perpendicular to the dispersion direction (see Sect.~\ref{obs-reduction}). We then
summed up the number of image columns corresponding to the actual slit widths, rebinned the resulting one-dimensional image to the actual pixel size, and fitted a Gaussian to derive the photocentre shift. The resulting shifts are compared with the range of shifts found for the blue components of the $^{13}$CO(1$-$0) lines in Fig.~\ref{shift-simulations}. Based on this comparison, the signatures for 2010 and 2013 can be reproduced by a bright blob at PA$\approx$0\degr, a distance between 70\,mas and 210\,mas, and a flux of more than 5\% and less than 20\% of the stellar flux (in the line centre). The parameter combinations that agree best with our data are 10\% flux ratio and 70 to 140\,mas distance. For a distance of 210\,mas, a flux ratio of 20\% is an acceptable fit to PAs up to 30\degr\ (Figs.~\ref{3mumeanshifts}, \ref{meanshifts}). In general, the simulations predict shifts that are too low for PA=60\degr, which indicates a larger extension of the blob than assumed. In the above parameter range,
the predicted increase in FWHM is similar to what is observed (Sect.~\ref{obs-reduction}). Larger distances and flux ratios can be excluded. These would show up as secondary peaks in the flux profiles along the slit, but no such structures were seen. 

Considering the above-mentioned measurements of the stellar radius, the distance range found for the blob corresponds to roughly 14 to 52 stellar radii, i.e.\ quite far away from the star but inwards of the blob-like structure observed by \citet{cruzalebes98}. The position angle is also different. We note at this point that the strength of the red-shifted absorption components shows no dependency on PA, indicating that its origin is within about 200\,mas, consistent with the likely blob distance.
Blob-like structures are found around red supergiants but typically at smaller distances and with a brightness of 1\% or less of the stellar brightness (e.g. \citealp{kervella2009, smithhinkleryde}). 

The photocentre shifts of the CO and HCN lines can also be used to assess the spectral energy distribution of the blob. The $^{13}$CO lines are considerably deeper than the HCN lines, and the continuum flux is also lower at 4.6\,$\mu$m than around 3\,$\mu$m, but the photocentre shifts found for these two sets of lines are basically identical. This implies a larger flux of the blob at 3\,$\mu$m. 
Interpolating the photospheric continuum fluxes at these wavelengths from the ISO-SWS spectrum (e.g.\ \citealp{klotz13}), using the typical depths of the blue line components shown in Fig.~\ref{shifts-depth}, and adopting the above contrast between blob and line flux of 5\% to 10\%, the ratio of the blob fluxes between 3\,$\mu$m and 4.6\,$\mu$m ranges from 2 to 8. The lower limit roughly corresponds to the continuum flux ratio of the ISO-SWS spectrum, while the Rayleigh-Jeans value of 5.5 is within the above range. 

Interpreting the flux ratio in terms of a temperature, the blob would have at least 1000\,K. We note at this point that the blob radiation cannot be scattered photospheric light as scattering is very inefficient at these long wavelengths. A rise in flux towards short wavelengths is also found for the blobs in the outer envelope of $\alpha$\,Ori \citep{kervella2009}. Thus the blob could be due to a mass ejection from the surface, although the brightness seems unusually high for its distance from the star as mentioned above. The red components of the absorption lines showing photocentre shifts can then be understood if they originate in cool, C-rich gas between the photosphere and the blob. The velocity difference between the blue and red components is of the order of 10\,km\,s$^{-1}$ 
and is close to the wind expansion velocity found from CO rotational lines \citep{schoierolofsson}. If this velocity is adopted as the expansion velocity of the blob and a distance of 275\,pc \citep{jorissen11} is used, the blob would have only moved by 22\,mas from 2010 to 2013, i.e.\ much less than the distance range predicted by our simulations. 
To find out whether the strength of the absorption and the similarity of the spectrum with that of the photosphere is compatible with the scenario of a blob in the wind would require elaborate simulations which are beyond the scope of this paper. 

The blob observed by \citet{cruzalebes98} and the blob detected by us could be the same mass ejection product at different positions of the ejection trajectory. However, the relative flux from the blob was smaller or equal in 1994, when the observations of \citet{cruzalebes98} were carried out, and it is still rather warm twenty years later. Such behaviour is not really expected.    

Since the scenario of a blob ejected from the star contains several uncertainties, it seems worthwhile to consider the possibility that the blob is a companion object. After adopting a minimum orbital radius of 70\,mas and a mass of the star between 2\,$M_{\sun}$ and 2.5\,$M_{\sun}$ \citep{klotz13}, the orbital periods (for a circular orbit) range between 40 and 50~years for companion masses between 2\,$M_{\sun}$ and 0.5\,$M_{\sun}$, respectively. Such long periods would be compatible with the similar shifts and line profile shapes found for 2010 and 2013. By neglecting the mass of the companion and assuming a maximum stellar mass of 2.5\,$M_{\sun}$, as well as an orbital radius of 70\,mas, the maximum circular orbital velocity is 10.8\,km\,s$^{-1}$, which agrees with the velocity difference found between the blue and red components and is slightly higher than the velocity variations found in \citet{jorissen11}. However, the flux ratios between any dwarf or substellar companion and the photosphere of TX Psc are much smaller than what we observe at 3\,$\mu$m and 4.6\,$\mu$m. A cool (sub)giant companion would have the right luminosity. For an equal age of both companions, the masses of the two components must be quite similar because TX~Psc is on the upper part of the AGB. If we attribute the blue and red line components to the primary and secondary, respectively, the radial velocity amplitudes expected for two components with masses between 1.5\,$M_{\sun}$ and 2.5\,$M_{\sun}$ range from 5\,km\,s$^{-1}$ to 9\,km\,s$^{-1}$ with the typical maximum velocity \textit{difference} being 14\,km\,s$^{-1}$. This is compatible with the observed velocity difference between the blue and red components, but the positions of the blue components of the $^{13}$CO$(1-0)$ lines show no significant blue shift relative to the systemic velocity. Furthermore, since the observed red HCN line components point to a carbon-rich chemistry of the gas responsible for the absorption, this excludes normal (sub)giants as companions, and the companion would have to be in the carbon star phase, too. Since TX~Psc apparently became a C star just recently \citep{klotz13}, we consider the scenario of two carbon stars orbiting each other rather unlikely. A possible solution could be the presence of accretion, although TX~Psc shows no other evidence for such a scenario. Therefore further observations are required to clarify the origin of the emission causing the photocentre shifts.

\begin{figure}
\centering
\includegraphics*[width=9.cm]{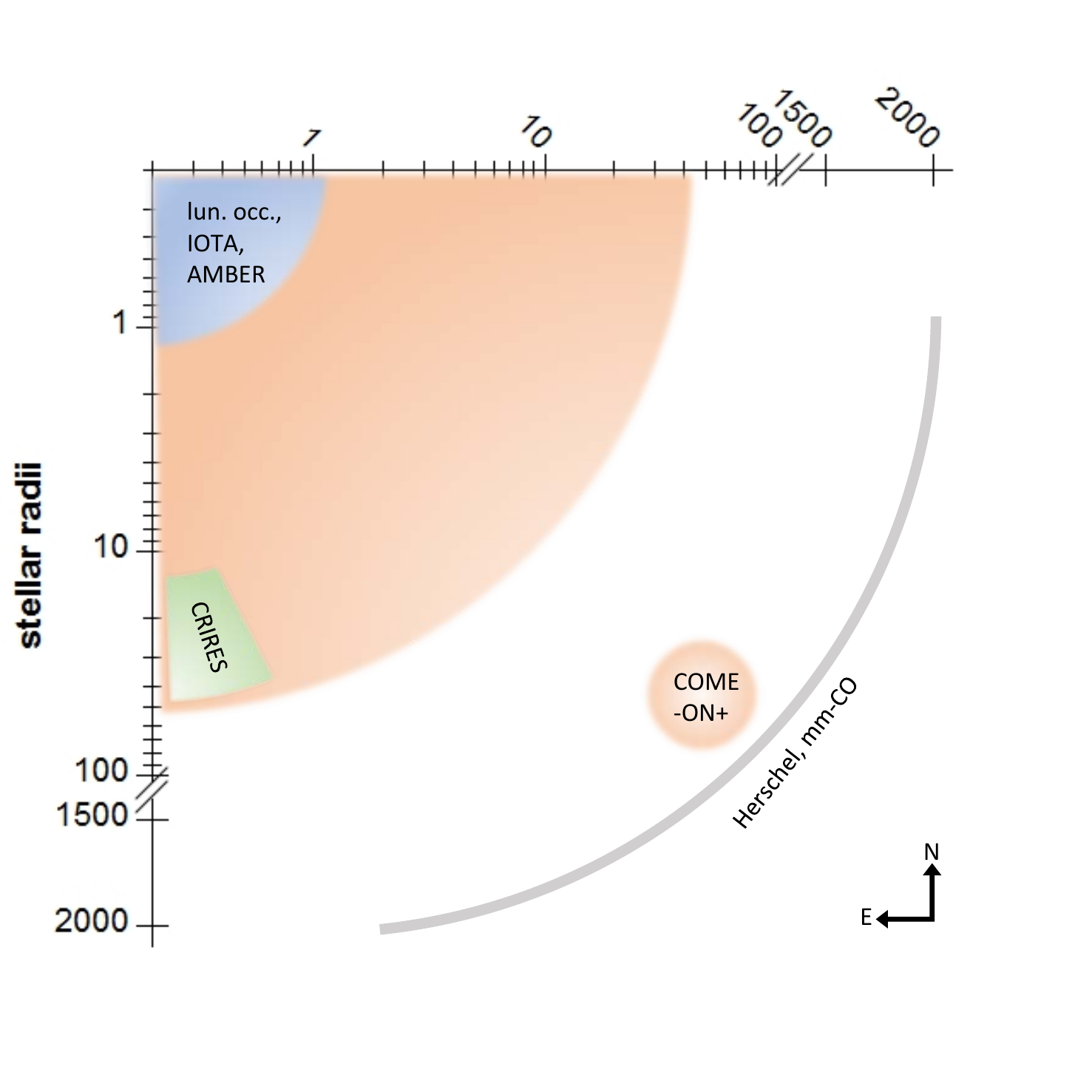}
\caption{\label{sketch} Sketch of the environment of TX~Psc (see also Sect.~\ref{target}). In the blue region, interferometric methods found asymmetries in various directions at scales up to the stellar radius. The orange regions show the structures observed by \citet{cruzalebes98}, the grey shell indicates the far-IR observations and the green region marks the blob detected in this work. Note the logarithmic scale! A stellar radius of 5\,mas was adopted.}
\end{figure}

\section{Summary and conclusion}
\label{conclusions}
We have presented the first spectro-astrometric study of a carbon-rich AGB star. The CRIRES observations of several CO $\Delta$v=1 lines near 4.6\,$\mu$m and HCN lines near 3\,$\mu$m showed significant photocentre shifts with a clear dependence on position angle. In all cases, tilde-shaped signatures were found where the positive and negative shifts (at PA~0\degr) are associated with blue and weaker red 
components of the lines. Observations in 2010 and 2013 showed very similar results. The shifts can be modelled with a bright spot 70\,mas to 210\,mas south of the star with a flux of several percent of the photospheric flux. The location of the blob is different from asymmetric structures found in previous observations, but these observations covered either larger or smaller angular scales. Combining the observations in the two different wavelength regions, we estimated a lower limit of the temperature of 1000\,K. The blob may be related to a mass ejection as found for AGB stars or red supergiants, although the flux associated with it is rather high, and it is not clear that such a blob located behind outflowing circumstellar material could produce the observed red absorption line components. We also considered the scenario of a companion object. While some of the blob properties are consistent with this, the observed flux is much too high. 

Figure~\ref{sketch} summarizes the structures detected in previous observations, together with the result of this work. The spectro-astrometry nicely closes the gap  between angular scales sampled by the interferometric data and the measurements using adaptive optics, i.e.\ the range of a few tens to a few hundred milli-arcseconds. However, there seems to be no general pattern or link between the different structures. This makes the interpretation of the structures in terms of mass ejections somehow more likely than
in terms of a companion. We therefore have to conclude that although there is clear spectro-astrometric evidence of a rather prominent structure near TX~Psc, no definite explanation can be given. Our data underline the very complex structure of the environment of this star, but further observations that sample the angular scales out to a few hundred milli-arcseconds are needed to get a clearer picture. While spectro-astrometry combined with adaptive optics can be considered as a very efficient tool for a first assessment of the geometry and kinematics of complex objects, imaging by optical interferometry (e.g.\ \citealp{berger12}) is a logical complement to this method. For cool stars, the upcoming MATISSE instrument \citep{matisse_short} offers a particularly high potential.

\begin{acknowledgements}
This work is supported by the Austrian Science Fund FWF under project numbers P23006-N16, P22911-N16, and P23737-N16. C.P.\ acknowledges  financial  support  from  the  FNRS
Research   Fellowship   –  Charg{\'e}   de   Recherche.
B.A.\ acknowledges the support from the {\em project STARKEY} funded by the ERC Consolidator Grant, G.A.\ n.~615604. A. Jorissen and the referee are thanked for careful reading of the manuscript and constructive comments. This research made use of the SIMBAD database, operated at the CDS, Strasbourg, France. 
\end{acknowledgements}

\bibliographystyle{aa} 
 \bibliography{hron} 

\begin{thebibliography}{36}
\expandafter\ifx\csname natexlab\endcsname\relax\def\natexlab#1{#1}\fi

\bibitem[{{Aoki} {et~al.}(1998){Aoki}, {Tsuji}, \& {Ohnaka}}]{aoki98b}
{Aoki}, W., {Tsuji}, T., \& {Ohnaka}, K. 1998, \aap, 333, L19

\bibitem[{{Aringer} {et~al.}(2009){Aringer}, {Girardi}, {Nowotny}, {Marigo}, \&
  {Lederer}}]{aringer09}
{Aringer}, B., {Girardi}, L., {Nowotny}, W., {Marigo}, P., \& {Lederer}, M.~T.
  2009, \aap, 503, 913

\bibitem[{{Bailey}(1998)}]{bailey98}
{Bailey}, J. 1998, \mnras, 301, 161

\bibitem[{{Berger} {et~al.}(2012){Berger}, {Malbet}, {Baron}, {Chiavassa},
  {Duvert}, {Elitzur}, {Freytag}, {Gueth}, {H{\"o}nig}, {Hron}, {Jang-Condell},
  {Le Bouquin}, {Monin}, {Monnier}, {Perrin}, {Plez}, {Ratzka}, {Renard},
  {Stefl}, {Thi{\'e}baut}, {Tristram}, {Verhoelst}, {Wolf}, \&
  {Young}}]{berger12}
{Berger}, J.-P., {Malbet}, F., {Baron}, F., {et~al.} 2012, \aapr, 20, 53

\bibitem[{{Brannigan} {et~al.}(2006){Brannigan}, {Takami}, {Chrysostomou}, \&
  {Bailey}}]{brannigan06}
{Brannigan}, E., {Takami}, M., {Chrysostomou}, A., \& {Bailey}, J. 2006,
  \mnras, 367, 315

\bibitem[{{Chiavassa} {et~al.}(2010){Chiavassa}, {Haubois}, {Young}, {Plez},
  {Josselin}, {Perrin}, \& {Freytag}}]{chiavassa10}
{Chiavassa}, A., {Haubois}, X., {Young}, J.~S., {et~al.} 2010, \aap, 515, A12

\bibitem[{{Cruzal{\`e}bes} {et~al.}(2015){Cruzal{\`e}bes}, {Jorissen},
  {Chiavassa}, {Paladini}, {Rabbia}, \& {Spang}}]{cruzalebes15}
{Cruzal{\`e}bes}, P., {Jorissen}, A., {Chiavassa}, A., {et~al.} 2015, \mnras,
  446, 3277

\bibitem[{{Cruzal{\`e}bes} {et~al.}(2013){Cruzal{\`e}bes}, {Jorissen},
  {Rabbia}, {Sacuto}, {Chiavassa}, {Pasquato}, {Plez}, {Eriksson}, {Spang}, \&
  {Chesneau}}]{cruzalebes13}
{Cruzal{\`e}bes}, P., {Jorissen}, A., {Rabbia}, Y., {et~al.} 2013, \mnras, 434,
  437

\bibitem[{{Cruzal{\`e}bes} {et~al.}(1998){Cruzal{\`e}bes}, {Lopez}, {Bester},
  {Gendron}, \& {Sams}}]{cruzalebes98}
{Cruzal{\`e}bes}, P., {Lopez}, B., {Bester}, M., {Gendron}, E., \& {Sams}, B.
  1998, \aap, 338, 132

\bibitem[{{Heske} {et~al.}(1989){Heske}, {te Lintel Hekkert}, \&
  {Maloney}}]{heske89b}
{Heske}, A., {te Lintel Hekkert}, P., \& {Maloney}, P.~R. 1989, \aap, 218, L5

\bibitem[{{Hinkle} {et~al.}(1995){Hinkle}, {Wallace}, \&
  {Livingston}}]{hinkleatlas95}
{Hinkle}, K., {Wallace}, L., \& {Livingston}, W.~C. 1995, {Infrared atlas of
  the Arcturus spectrum, 0.9-5.3 microns} (San Francisco, Calif.~: Astronomical
  Society of the Pacific, 1995.)

\bibitem[{{Hron} {et~al.}(1998){Hron}, {Loidl}, {H{\"o}fner}, {Jorgensen},
  {Aringer}, \& {Kerschbaum}}]{hron_rscl}
{Hron}, J., {Loidl}, R., {H{\"o}fner}, S., {et~al.} 1998, \aap, 335, L69

\bibitem[{{J{\o}rgensen} {et~al.}(2000){J{\o}rgensen}, {Hron}, \&
  {Loidl}}]{jorgensen00}
{J{\o}rgensen}, U.~G., {Hron}, J., \& {Loidl}, R. 2000, \aap, 356, 253

\bibitem[{{J{\o}rgensen} \& {Johnson}(1991)}]{jorgensen91}
{J{\o}rgensen}, U.~G. \& {Johnson}, H.~R. 1991, \aap, 244, 462

\bibitem[{{Jorissen} {et~al.}(2011){Jorissen}, {Mayer}, {Van Eck},
  {Ottensamer}, {Kerschbaum}, {Ueta}, {Bergman}, {Blommaert}, {Decin},
  {Groenewegen}, {Hron}, {Nowotny}, {Olofsson}, {Posch}, {Sjouwerman},
  {Vandenbussche}, \& {Waelkens}}]{jorissen11}
{Jorissen}, A., {Mayer}, A., {Van Eck}, S., {et~al.} 2011, \aap, 532, A135

\bibitem[{{K{\"a}ufl} {et~al.}(2004){K{\"a}ufl}, {Ballester}, {Biereichel},
  {Delabre}, {Donaldson}, {Dorn}, {Fedrigo}, {Finger}, {Fischer}, {Franza},
  {Gojak}, {Huster}, {Jung}, {Lizon}, {Mehrgan}, {Meyer}, {Moorwood}, {Pirard},
  {Paufique}, {Pozna}, {Siebenmorgen}, {Silber}, {Stegmeier}, \&
  {Wegerer}}]{kaeuflcriresspie}
{K{\"a}ufl}, H.-U., {Ballester}, P., {Biereichel}, P., {et~al.} 2004, in
  Society of Photo-Optical Instrumentation Engineers (SPIE) Conference Series,
  Vol. 5492, Ground-based Instrumentation for Astronomy, ed. A.~F.~M.
  {Moorwood} \& M.~{Iye}, 1218--1227

\bibitem[{{Kervella} {et~al.}(2009){Kervella}, {Verhoelst}, {Ridgway},
  {Perrin}, {Lacour}, {Cami}, \& {Haubois}}]{kervella2009}
{Kervella}, P., {Verhoelst}, T., {Ridgway}, S.~T., {et~al.} 2009, \aap, 504,
  115

\bibitem[{{Klotz} {et~al.}(2013){Klotz}, {Paladini}, {Hron}, {Aringer},
  {Sacuto}, {Marigo}, \& {Verhoelst}}]{klotz13}
{Klotz}, D., {Paladini}, C., {Hron}, J., {et~al.} 2013, \aap, 550, A86

\bibitem[{{Le Bouquin} {et~al.}(2009){Le Bouquin}, {Lacour}, {Renard},
  {Thi{\'e}baut}, {Merand}, \& {Verhoelst}}]{lebouquin09}
{Le Bouquin}, J.-B., {Lacour}, S., {Renard}, S., {et~al.} 2009, \aap, 496, L1

\bibitem[{{Lesage} \& {Wiedemann}(2014)}]{lesage2014}
{Lesage}, A.-L. \& {Wiedemann}, G. 2014, \aap, 563, A86

\bibitem[{{Lopez} {et~al.}(2012){Lopez}, {Lagarde}, {Antonelli},
  {et~al.}}]{matisse_short}
{Lopez}, B., {Lagarde}, S., {Antonelli}, P., {et~al.} 2012, in Society of
  Photo-Optical Instrumentation Engineers (SPIE) Conference Series, Vol. 8445,
  Society of Photo-Optical Instrumentation Engineers (SPIE) Conference Series

\bibitem[{{Luttermoser}(2000)}]{luttermoser00}
{Luttermoser}, D.~G. 2000, in IAU Symposium, Vol. 177, The Carbon Star
  Phenomenon, ed. R.~F. {Wing}, 105

\bibitem[{{Maercker} {et~al.}(2012){Maercker}, {Mohamed}, {Vlemmings},
  {Ramstedt}, {Groenewegen}, {Humphreys}, {Kerschbaum}, {Lindqvist},
  {Olofsson}, {Paladini}, {Wittkowski}, {de Gregorio-Monsalvo}, \&
  {Nyman}}]{maerckernature}
{Maercker}, M., {Mohamed}, S., {Vlemmings}, W.~H.~T., {et~al.} 2012, \nat, 490,
  232

\bibitem[{{Ohnaka} {et~al.}(2013){Ohnaka}, {Hofmann}, {Schertl}, {Weigelt},
  {Baffa}, {Chelli}, {Petrov}, \& {Robbe-Dubois}}]{ohnaka13antares}
{Ohnaka}, K., {Hofmann}, K.-H., {Schertl}, D., {et~al.} 2013, \aap, 555, A24

\bibitem[{{Olofsson} {et~al.}(1993){Olofsson}, {Eriksson}, {Gustafsson}, \&
  {Carlstr{\"o}m}}]{olofsson93}
{Olofsson}, H., {Eriksson}, K., {Gustafsson}, B., \& {Carlstr{\"o}m}, U. 1993,
  \apjs, 87, 305

\bibitem[{{Paladini} {et~al.}(2012){Paladini}, {Sacuto}, {Klotz}, {Ohnaka},
  {Wittkowski}, {Nowotny}, {Jorissen}, \& {Hron}}]{paladini12}
{Paladini}, C., {Sacuto}, S., {Klotz}, D., {et~al.} 2012, \aap, 544, L5

\bibitem[{{Pontoppidan} {et~al.}(2008){Pontoppidan}, {Blake}, {van Dishoeck},
  {Smette}, {Ireland}, \& {Brown}}]{pontoppidan08}
{Pontoppidan}, K.~M., {Blake}, G.~A., {van Dishoeck}, E.~F., {et~al.} 2008,
  \apj, 684, 1323

\bibitem[{{Ragland} {et~al.}(2006){Ragland}, {Traub}, {Berger}, {Danchi},
  {Monnier}, {Willson}, {Carleton}, {Lacasse}, {Millan-Gabet}, {Pedretti},
  {Schloerb}, {Cotton}, {Townes}, {Brewer}, {Haguenauer}, {Kern}, {Labeye},
  {Malbet}, {Malin}, {Pearlman}, {Perraut}, {Souccar}, \&
  {Wallace}}]{ragland06}
{Ragland}, S., {Traub}, W.~A., {Berger}, J.-P., {et~al.} 2006, \apj, 652, 650

\bibitem[{{Richichi} {et~al.}(1995){Richichi}, {Chandrasekhar}, {Lisi},
  {Howell}, {Meyer}, {Rabbia}, {Ragland}, \& {Ashok}}]{richichi95}
{Richichi}, A., {Chandrasekhar}, T., {Lisi}, F., {et~al.} 1995, \aap, 301, 439

\bibitem[{{Ridgway} {et~al.}(1978){Ridgway}, {Carbon}, \& {Hall}}]{ridgway78}
{Ridgway}, S.~T., {Carbon}, D.~F., \& {Hall}, D.~N.~B. 1978, \apj, 225, 138

\bibitem[{{Ryde} {et~al.}(1999){Ryde}, {Gustafsson}, {Hinkle}, {Eriksson},
  {Lambert}, \& {Olofsson}}]{ryde99}
{Ryde}, N., {Gustafsson}, B., {Hinkle}, K.~H., {et~al.} 1999, \aap, 347, L35

\bibitem[{{Samus} {et~al.}(2009){Samus}, {Durlevich}, \& {et al.}}]{samus09}
{Samus}, N.~N., {Durlevich}, O.~V., \& {et al.} 2009, VizieR Online Data
  Catalog, 1, 2025

\bibitem[{{Sch{\"o}ier} \& {Olofsson}(2001)}]{schoierolofsson}
{Sch{\"o}ier}, F.~L. \& {Olofsson}, H. 2001, \aap, 368, 969

\bibitem[{{Smith} {et~al.}(2009){Smith}, {Hinkle}, \& {Ryde}}]{smithhinkleryde}
{Smith}, N., {Hinkle}, K.~H., \& {Ryde}, N. 2009, \aj, 137, 3558

\bibitem[{{Voigt}(2009)}]{voigt09a}
{Voigt}, B. 2009, PhD thesis, Hamburg University, Germany

\bibitem[{{Voigt} \& {Wiedemann}(2009)}]{voigt09b}
{Voigt}, B. \& {Wiedemann}, G. 2009, in American Institute of Physics
  Conference Series, Vol. 1094, 15th Cambridge Workshop on Cool Stars, Stellar
  Systems, and the Sun, ed. E.~{Stempels}, 896--899

\end{thebibliography}

\Online

 \begin{appendix}
\section{Further flux and position spectra for CO and HCN} 
 \begin{figure}[h]
\centering
\includegraphics*[width=9.cm]{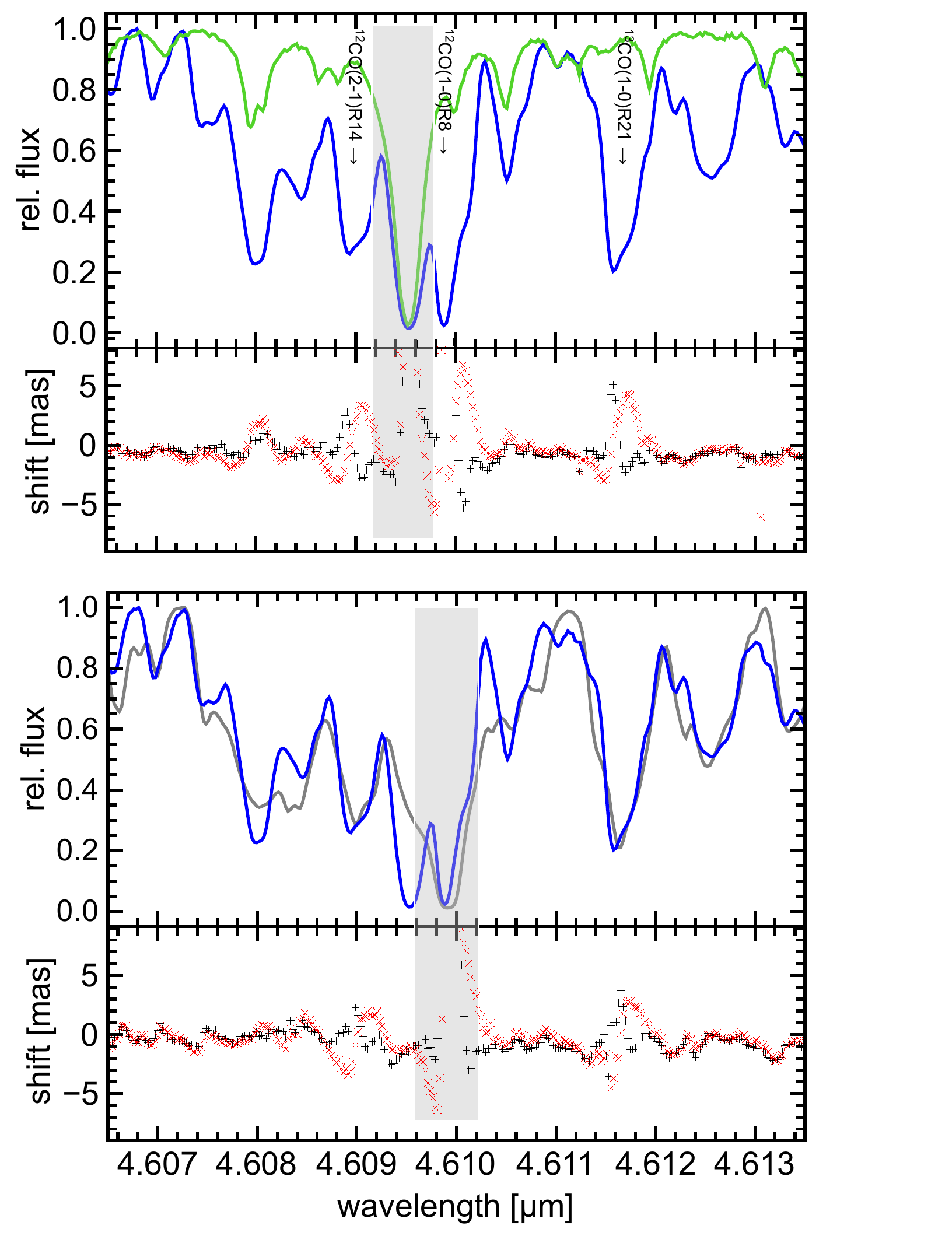}
\caption{\label{posspectra2} Same as Fig.~\ref{posspectra1} but for the region near 4.61\,$\mu$m.}
\end{figure}
\begin{figure}
\centering
\includegraphics*[width=9.cm]{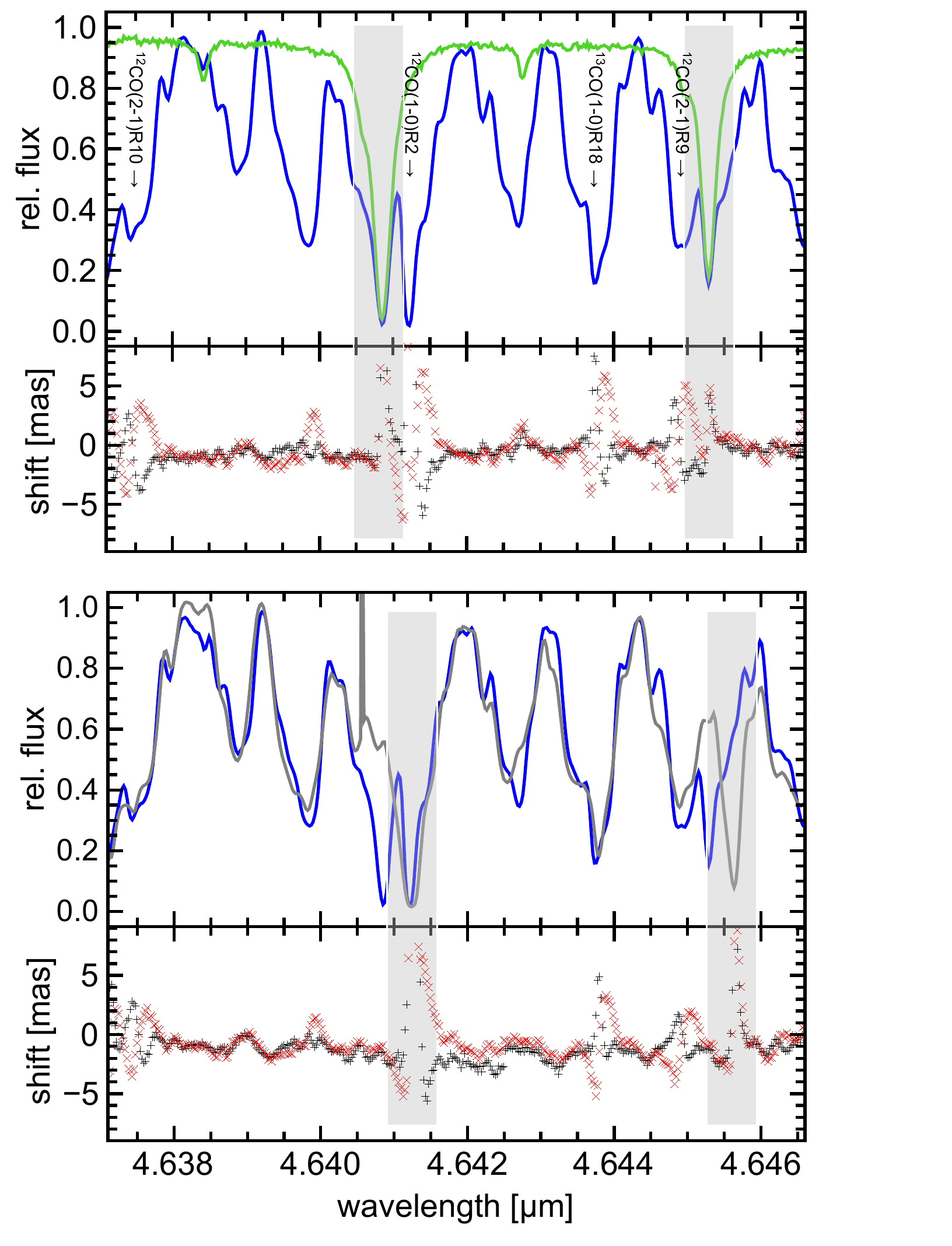}
\caption{\label{posspectra3} Same as Fig.~\ref{posspectra1} but for the region near 4.64\,$\mu$m.}
\end{figure}

\begin{figure}
\centering
\includegraphics*[width=9.cm]{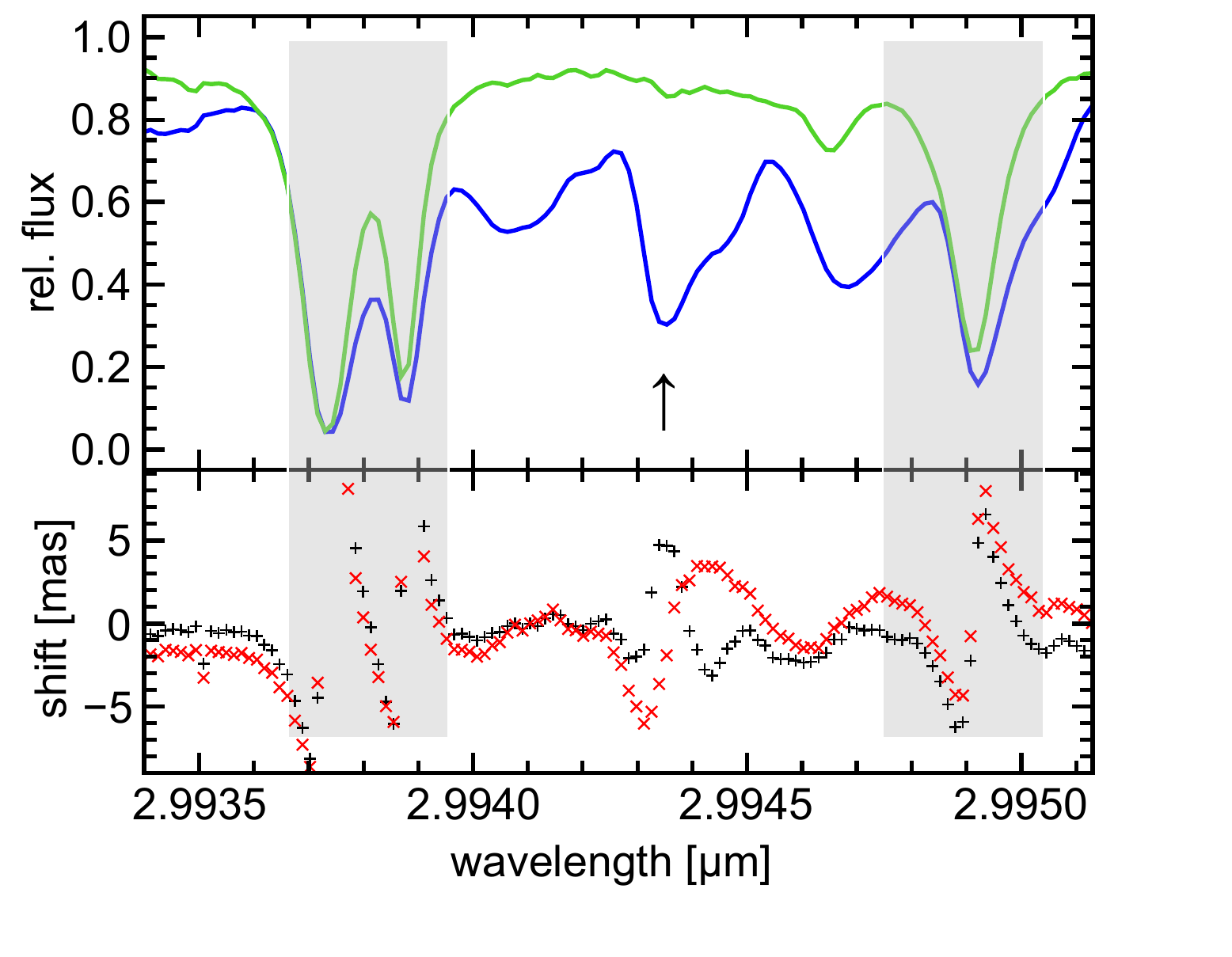}
\caption{\label{3muposspectra2} Same as Fig.~\ref{3muposspectra1} but for the region near 2.99\,$\mu$m.}
\end{figure}

\begin{figure}
\centering
\includegraphics*[width=9.cm]{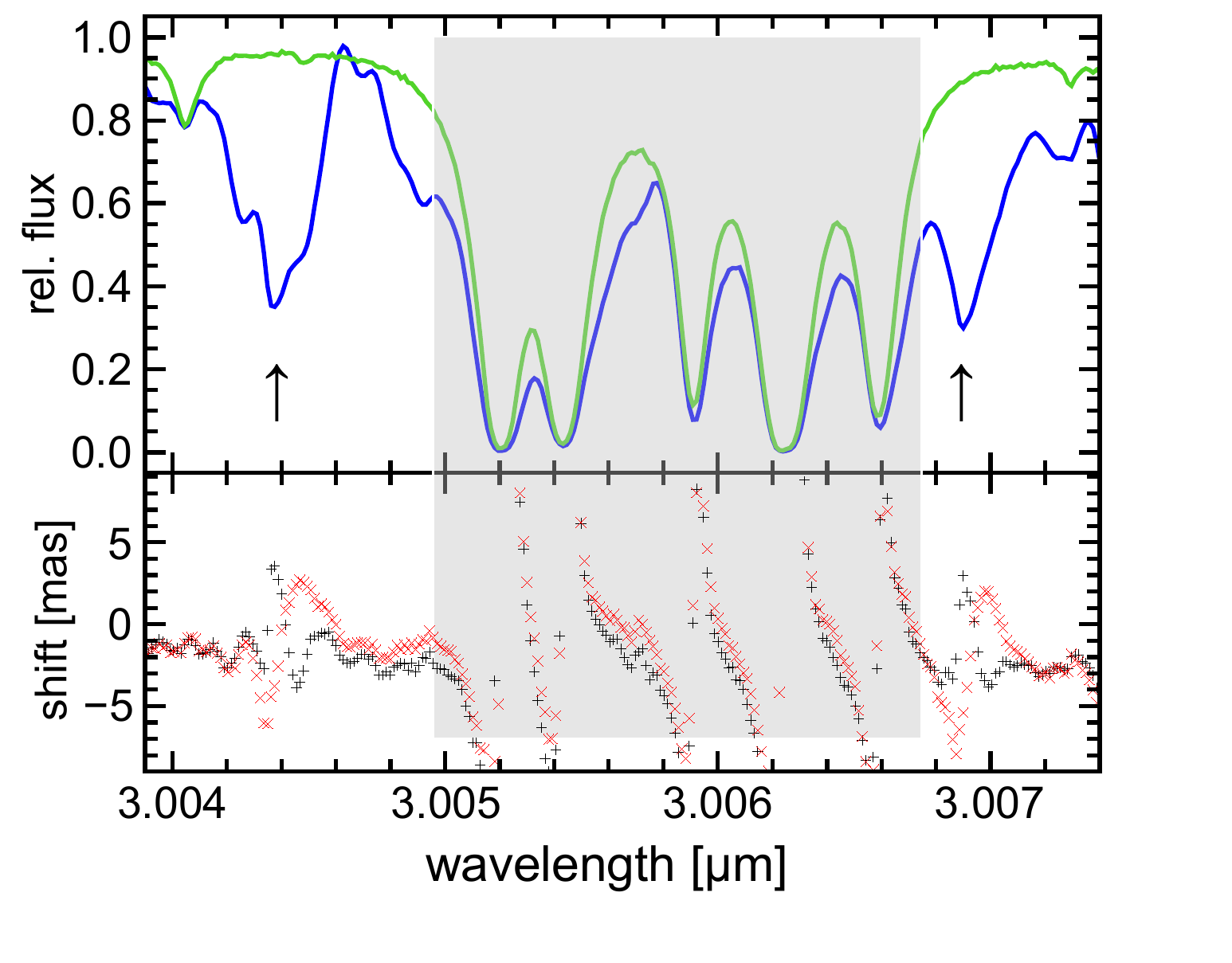}
\caption{\label{3muposspectra3} Same as Fig.~\ref{3muposspectra1} but for the region near 3.00\,$\mu$m.}
\end{figure}

\end{appendix}

\end{document}